\documentclass[11pt]{article}
\usepackage[utf8]{inputenc}
\usepackage[margin=0.7in]{geometry}
\usepackage{amsfonts, amsmath}

\usepackage{array}
\usepackage{graphicx}
\usepackage{svg}
\usepackage{setspace}

\usepackage[round,comma,authoryear]{natbib}
\usepackage{multibib}
\newcites{app}{Supplementary references}

\title{\huge \textbf{Balancing the cellular budget: lessons in metabolism from microbes to cancer}}
\date{\empty}
\usepackage{authblk}
\author[1]{B. Vibishan}
\author[1,*]{Mohit Kumar Jolly}
\author[2,*]{Akshit Goyal}
\affil[1]{Department of Bioengineering, Indian Institute of Science (IISc), Bengaluru, India}
\affil[2]{International Centre for Theoretical Sciences (ICTS), Tata Institute of Fundamental Research (TIFR), Bengaluru, India}
\affil[*]{Corresponding authors: akshitg@icts.res.in, mkjolly@iisc.ac.in}

\usepackage{hyperref}
\hypersetup{
	colorlinks = true, 
	urlcolor = green, 
	linkcolor = blue, 
	citecolor = red 
}
\begin{document}

\pagenumbering{gobble}
\maketitle

\newpage
\pagenumbering{roman}
\section*{\centering Abstract}
\paragraph{\empty}Cancer cells are often seen to prefer glycolytic metabolism over oxidative phosphorylation even in the presence of oxygen-a phenomenon termed the Warburg effect. Despite significant strides in the decades since its discovery, a clear basis is yet to be established for the Warburg effect and why cancer cells show such a preference for aerobic glycolysis. In this review, we draw on what is known about similar metabolic shifts both in normal mammalian physiology and overflow metabolism in microbes to shed new light on whether aerobic glycolysis in cancer represents some form of optimisation of cellular metabolism. From microbes to cancer, we find that metabolic shifts favouring glycolysis are sometimes driven by the need for faster growth, but the growth rate is by no means a universal goal of optimal metabolism. Instead, optimisation goals at the cellular level are often multi-faceted and any given metabolic state must be considered in the context of both its energetic costs and benefits over a range of environmental contexts. For this purpose, we identify the conceptual framework of resource allocation as a potential testbed for the investigation of the cost-benefit balance of cellular metabolic strategies. Such a framework is also readily integrated with dynamical systems modelling, making it a promising avenue for new answers to the age-old question of why cells, from cancers to microbes, choose the metabolic strategies they do.

\paragraph{Keywords:}Warburg effect; microbial metabolism; bioenergetics; phenotypic heterogeneity; resource allocation
\paragraph{Word count:}5662

\newpage
\section*{Author contributions}
All authors conceptualised the review. BV and AG collected and reviewed the primary literature, and BV wrote the first draft. All authors contributed to analysis and critical revisions, and approved the final version.

\section*{Funding statement}
BV and MKJ were supported by the Param Hansa Philanthropies. AG acknowledges support from the Ashok and Gita Vaish Junior Researcher Award, the Department of Science and Technology-Science and Engineering Research Board (DST-SERB) Ramanujan Fellowship, as well as the Department of Atomic Energy (DAE), Govt of India, under project no. RTI4001.

\section*{Competing interests}
The authors declare no conflicts of interest.

\newpage
\pagenumbering{arabic}
\section{Introduction}

\paragraph{\empty}In 1956, Otto Warburg found that pure cultures of cancer cells showed remarkably high rates of fermentation in the presence of oxygen \citep{warburg1956a}, kickstarting a decades-long investigation of how cancer cells support their relentless growth. In the intervening years, a great deal has been discovered regarding the apparent switch made by cancer cells from mitochondrial oxidative phosphorylation to aerobic fermentation as the source of ATP at high growth rates. Indeed, the current understanding of cancer metabolism is undoubtedly more richly detailed than what was known at the time of Warburg’s discovery \citep{mayers_famine_2015}. While initial hypotheses to explain this effect in cancer focused on a defective mitochondrial apparatus \citep{warburg1956a}, later work showed that mitochondrial function remained active in several cancers \citep{weinhouse1967, jose2011, potter2016}. Nevertheless, some fundamental questions remain regarding the nature of metabolism-related decision making in cancer systems.

\paragraph{\empty}The Warburg effect has occasionally been taken to mean that aerobic fermentation is largely in opposition to mitochondrial oxidative phosphorylation, such that the cancer cell chooses between two mutually-exclusive metabolic strategies \citep{molenaar2009a, hsu_cancer_2008}. However, recent work is beginning to question this supposed dichotomy, advocating for a more holistic view of how and why cancer cells manage the balance of fermentation and oxidative phosphorylation \citep{jacquet_searching_2022, jacquet2023}. Detailed mathematical models accounting for the stoichiometry of ATP production and consumption through different metabolic pathways in cancer cells have likewise advocated for re-evaluating the conventional notion of a “switch” between oxidative and fermentative metabolism \citep{nath2024}. It is important to resolve these positions, simply because fermentation and oxidative phosphorylation differ widely in their energetic output and biosynthetic roles \citep{nelson2017, nath2024}. A little probing in this context reveals a large web of opinions, observations, methodologies and hypotheses addressing the nature of aerobic fermentation in cancer, although a broad consensus is lacking. In this review, we attempt to draw a more comprehensive view of the various findings and theories surrounding cancer metabolism, identifying common threads wherever possible to elucidate the current understanding of fermentation in cancer cells.

\paragraph{\empty}In a broader sense, the phenomenon of aerobic fermentation in cancer also seems to parallel overflow metabolism in bacteria and some single-celled eukaryotes, both of which show a clear transition away from oxidative phosphorylation or other high-ATP yield metabolic pathways in favour of fermentative growth when substrates are abundant \citep{meyer1984, wolfe2005, vemuri2006, molenaar2009a, peebo2015, basan2015}. As with cancer, a wide range of explanations have been proposed and tested for this phenomenon over the years \citep{vemuri2006, scott2010, schuetz2012, you2013, new2014, towbin2017}, but one prominent line of work has developed an understanding of microbial metabolic strategies in terms of their cost-vs-benefit to the cell \citep{molenaar2009a, metzl-raz2017, chure2023}. This framework has been termed resource allocation \citep{basan2015}, and its application has provided a new perspective on how unicellular organisms regulate their metabolic requirements that goes beyond a simple rate-vs-yield tradeoff over ATP production. We will therefore provide an overview of the main findings from this literature with the aim of illustrating how the approach works for microbial systems. We will then highlight the various parallels in metabolism in microbes and cancer, which illustrate the potential utility of resource allocation theory as a rational basis for cancer metabolism.
	
\paragraph{\empty}It would be useful here to provide a quick explanation of oxidative phosphorylation and fermentation, both to set the context and to clarify the terminology used. What we have referred to as aerobic fermentation has also been termed in various contexts as overflow metabolism, the Warburg effect \citep{vanderheiden2009}, aerobic glycolysis \citep{deberardinis2007}, the Crabtree effect \cite{dedeken1966}, and possibly other terms. All these refer to the cytosolic catabolism of glucose into pyruvate, primarily through the Embden-Meyerhof-Parnas (EMP) pathway (\citet{nelson2017}, but also see the Entner-Doudoroff pathway as in \citet{flamholz2013}), followed by the conversion of pyruvate to acetate, lactate or ethanol, finally producing up to 2 ATPs for every molecule of glucose. The process is usually accompanied by the secretion of either acetate \citep{meyer1984, vemuri2006}, ethanol \citep{vandijken1993}, or lactate \citep{mckeehan1982, hsu_cancer_2008}, and on a per mole basis, has a far lower ATP yield than oxidative phosphorylation \citep{nelson2017}. This latter pathway, often shortened to OXPHOS, shunts pyruvate into the tri-carboxylic acid (TCA) cycle or the Krebs cycle, eventually replenishing several reducing co-factors like NADH that enter the electron-transport chain (ETC) to produce up to 32 ATPs per molecule of glucose, accompanied by the reduction of molecular oxygen to water. In eukaryotes, glycolysis occurs in the cytoplasm whereas the Krebs cycle and the ETC are localised in the mitochondria \citep{nelson2017}. Over the course of this review, we will occasionally use the terms “glycolytic” and “oxidative” to describe cell states that predominantly run glycolysis and OXPHOS respectively.

\section{Glycolysis in mammalian cell metabolism}
\paragraph{\empty}The question of why cancer cells use aerobic glycolysis largely revolves around the fact that cancer cells grow much faster than non-cancerous cells. But changes in the net glycolytic flux occur in normal physiology in several contexts \citep{heiden2011}, and such shifts illustrate effectively that the regulation of glycolytic flux does not follow a simple correspondence with faster proliferation. Instead, they reveal a more nuanced balance between fast growth and biomass production that determines the overall metabolic phenotype of the cell. Importantly, this balance is sometimes centered on optimising glycolysis for net ATP yield, but in other cases, couples glycolysis with aspects of anabolism other than ATP production. We will review three illustrative examples here to understand the basis of these shifts in non-cancerous metabolism: (1) vascular endothelial cell (EC) metabolism during angiogenesis, (2) immune cell activation, and (3) intestinal crypt homeostasis. More such parallels might undoubtedly be found upon further investigation.

\paragraph{\empty}Metabolic shifts that occur during angiogenesis i.e., the formation of new blood vessels, depend on different demands of vascular endothelial cells (ECs) based on their position. Vascular ECs line the walls of blood vessels and are responsible for the sprouting and formation of new blood vessels in response to oxygen or nutrient deprivation signals in the form of the signaling protein, vascular endothelial growth factor (VEGF) secreted by stromal cells under deprivation \citep{debock2013, fitzgerald2018}. In response to a VEGF gradient, some ECs adopt a so-called tip phenotype and form the leading edge of the newly forming blood vessel. Tip cells are characterised by increased expression of the DLL4 ligand as well as increased glycolytic flux, extension of the cell membrane to form filopodia and increased cell migration along the increasing VEGF gradient \citep{debock2013, debock2013a}. This is also accompanied by the concomitant emergence of so-called stalk cells that form the body of the blood vessel. Notch1 signalling in stalk cells activated by DLL4 expressed by the tip cells inhibits VEGFR signalling in the stalk cells and prevents them from gaining tip cell characteristics \citep{benedito2009}. Interestingly, stalk cells prioritise biosynthesis by increased fatty acid metabolism and amino acid biogenesis while still remaining highly glycolytic. On the other hand, glycolysis addiction in tip cells is thought to stem from one of several reasons, including allowing for rapid metabolic remodelling \citep{pfeiffer2001}, increased availability of a range of precursors for amino acid and/or nucleotide biosynthesis \citep{pan2009, vandekeere2018}, higher competitive ability in the context of invading a new microenvironment \citep{pfeiffer2001}, and protection from oxidative stress \citep{debock2013}. 
	
\paragraph{\empty}T-lymphocytes, upon activation, become proliferative and this has been associated with increased glycolysis \textit{in vitro} \citep{marelli-berg2012, maciver2013}. In the context of cancer specifically, CD11b$^+$ myeloid cells and tumour-infiltrating CD4/8$^+$ T-cells have been shown to consume a higher amount of the available glucose than surrounding cancer cells which in turn are dependent on glutamine \citep{reinfeld_cell_2021}. Glucose limitation and reduced glycolysis has also been shown to lead to reduced T-cell activity in the tumour site and impaired tumour control \citep{chang2015}. As with the endothelial cells, this might suggest a direct association between glycolysis and higher cell proliferation and/or migration. Nevertheless, other studies have shown that such a metabolic shift might be an artefact of in vitro conditions, as activated T-cells in vivo seem to show a more dynamic immune metabolic phenotype that is centered on serine biosynthesis \citep{ma2019}. Interestingly, it has been shown that $\alpha$-ketoglutarate ($\alpha$-KG), a key Krebs cycle intermediate, inhibits M1 polarisation of macrophages upon treatment with interleukin-4 (IL-4) or lipo-polysaccharide (LPS) exposure in favour of the anti-inflammatory M2 phenotype \citep{liu2017}. In this case, however, a direct substantial connection does not exist between cell proliferation and glycolysis, so it is unclear if this role for $\alpha$-KG could also tie into a broader metabolic strategy dictating the choice between M1 and M2 polarisation.
	
\paragraph{\empty}It is well-known that intestinal crypts are organised into a continuum of stem to differentiated cell states, with the Lgr5$^+$ stem cells at the base of the crypt leading to progressively more differentiated cells that occupy higher positions along the crypt, up to the fully differentiated absorptive or secretory enterocytes at the top of each villus \citep{crompton1998, snippert2010}. It is has been shown that while the Lgr5$^+$ stem cells have high mitochondrial activity, the supporting Paneth cells are highly glycolytic and act as lactate sources to sustain mitochondrial metabolism in the remaining cells of the intestinal crypt \citep{rodriguez-colman2017}. Some evidence also suggests that intestinal stem cells are themselves also glycolytic rather than oxidative \citep{stringari2012}. Interestingly however, preventing the shunting of pyruvate into the Krebs cycle by inhibiting its transport into the mitochondria is sufficient to maintain the proliferation of intestinal Lgr5$^+$ stem cells \citep{schell2017}, whereas loss of the mitochondrial chaperone, HSP60 leads to a loss of stemness and cell proliferation \citep{berger2016}. These findings on the relevance of mitochondrial function for stem cell integrity and organisation in the intestinal crypts are seemingly contradictory, although it is possible that this relevance is itself context-dependent, based on whether there is ongoing renewal in the stem cell compartment. Nevertheless, glycolysis might broadly be related to stemness and proliferation, as shown for hematopoietic \citep{simsek2010} and a range of other stem cells \citep[\empty][and references therein]{ito2014}.
	
\paragraph{\empty}On the whole, non-cancerous metabolism offers several potential contexts in which a glycolytic state may be preferred over an oxidative state. Cell proliferation often seems to be a good predictor of a glycolytic state, with more glycolytic flux supporting more proliferation. This is also related to the fact that cytosolic glycolysis connects with a wide range of biosynthetic precursor pathways, including the pentose-phosphate shunt, one-carbon metabolism for serine biosynthesis, and hexosamine biosynthesis pathways. Pathways like the pentose-phosphate shunt also replenish co-factors like NADPH that can be used in other anabolic processes.Taken together, this suggests that a preference for glycolysis in normal physiology, either in the presence of oxygen (vascular ECs), or under hypoxia (some stem cells), could be broadly aligned with a strategy for higher proliferation that utilises glucose for other anabolic purposes rather than energy production alone. Of course, the latter case does not strictly require evoking a role for proliferation, since hypoxia should be expected to favour glycolysis under any circumstances. To some extent, glycolysis is also seen to mitigate oxidative stress, which is an inevitable byproduct of OXPHOS. Taken together, proliferation, hypoxia, and oxidative stress represent clear bases on which the balance between OXPHOS and glycolysis depends for normal physiology. 

\paragraph{\empty}Here, it is important to note that in all of these cases, cell metabolism seems to lie downstream of the cell state decision, such that other aspects of cell physiology or external factors determine what state the cell should adopt, following which a metabolic strategy is devised that facilitates this state. Such a situation in normal cellular metabolism may also be consistent with cancer metabolism, since somatic mutations or other regulatory changes in cancer cell physiology may lie upstream of metabolic pathways and dictate the metabolic strategy of the cell to align with a given genotype and phenotype.

\section{Cancer's context: how ``normal" is cancer metabolism?}
\paragraph{\empty}Some of the earliest and best-developed hypotheses regarding metabolic strategies in cancer attributed the relative advantage of aerobic glycolysis over OXPHOS to the fact that actively-dividing cells anywhere, cancerous or non-cancerous, must prioritise building biomass over high-yield energy production \citep{molenaar2009a, vanderheiden2009}. Indeed, much of the foregoing discussion on normal physiology would seem to support this notion. However, it would be simplistic to reduce cancer metabolism to the rate-vs-yield balance for OXPHOS and glycolysis. Many metabolic phenotypes have been identified in cancer that seem to serve a variety of other functional goals for the cell, some of which we will review here. These provide a broader perspective of cancer metabolism, and illustrate the need for a more integrative framework that can reconcile these diverse observations.

\paragraph{\empty}We have hinted above that the choice between OXPHOS and glycolysis is largely a matter of the rate of ATP production vs the yield of ATP per molecule of glucose. However, experimental evidence from normal and cancer cell lines suggests that OXPHOS and glycolysis may actually be serving distinct cellular needs, responding to perturbations over different timescales-OXPHOS, being the slower pathway, delivers a baseline energetic flux with high efficiency, while glycolysis responds rapidly to fluctuating energetic demands \citep{epstein2014}. In such a case then, it may not be strictly necessary for glycolysis to be preferred in all cancerous cells, as has been indicated elsewhere \citep{jose2011, potter2016, hensley_metabolic_2016}. It has also been shown in respiration-deficient mammalian cells that external supply of aspartate entirely rescues the loss of cell growth due to OXPHOS deficiency, which indicates a clear biosynthetic role for OXPHOS apart from ATP production \citep{Sullivan2015}. 

\paragraph{\empty}Another key aspect of cancer metabolism concerns glutamine catabolism. It is well-known that cancer cells are voracious consumers of the amino acid, glutamine \citep{deberardinis2007}. Interestingly, this consumption of glutamine is not linked to an increased demand for nitrogen, and it has been shown that glutamine acts primarily as a carbon source to support continued turning of the Krebs cycle \citep{deberardinis2007, fan2013, liu2017}. In this case therefore, cancer cells seem to use glucose for ATP production through glycolysis while glutamine is used to power biomass production by replenishing Krebs cycle intermediates. The other amino acid that has also been implicated in cancer growth is glycine \citep{jain_metabolite_2012}, although there is some debate in the field about whether it is glycine or serine that is the more important regulator of cancer cell growth \citep{maddocks2013, labuschagne2014}. Serine forms a part of one-carbon metabolism which links it to nucleotide synthesis \citep{labuschagne2014}, while glycine has been shown to be incorporated into DNA directly during cancer cell growth \citep{jain_metabolite_2012}. It is therefore unclear exactly where serine and glycine must be placed in the context of cancer metabolism, although both likely meet some biosynthetic demand.

\paragraph{\empty}Cancer metabolism should also be seen in the context of more than individual metabolites or metabolic pathways. In fact, it has been argued that the exact advantage of a given metabolic strategy must be determined by some combination of the tumour’s cell-of-origin, existing genetic lesions and/or somatic evolution, and the tumour environment \citep{luengo_targeting_2017}. Cell-of-origin and/or tissue-type effects have been documented to some degree in biopsies of cancer vs normal tissue \citep{elser_biological_2007} and in cell lines across different cancer types \citep{reznik_landscape_2018}. Clear evidence from mice studies also confirms that Myc-induced vs Met-induced lung tumours show very different metabolic characteristics, with Myc-induced lung tumours showing both higher glucose uptake and lactate secretion relative to surrounding normal tissue than Met-induced tumours \citep{yuneva_metabolic_2012}. This indicates that pre-existing genetic background has a direct effect on the observed metabolic states in the tumour \citep{brunner2023}. Likewise, considerable evidence also shows that the external environment has a significant effect on tumour metabolic strategies, with metabolic states responding to changes in availabilities of specific small metabolites like glucose, glutamine or cystine \citep{davidson2016, muir2017}, or broader differences in medium composition \citep{cantor2017, brunner2023}.

\paragraph{\empty}Altogether, these observations show that while some aspects of cancer metabolism can be explained in terms of normal physiological responses, many features of cancer metabolism require distinct explanations that are not rooted in normal mammalian physiology. In this context, we suggest that the current understanding of microbial metabolism, specifically of the factors underlying the phenomenon of overflow metabolism, offers potential clues for how these various aspects of cancer metabolism may be understood together. In the following section, we will offer a brief overview of overflow metabolism in microbes and what is understood about its mechanistic underpinnings. This will provide leads that we will then extend to cancer metabolism.

\section{Resource allocation: a rational basis for microbial metabolism}
\paragraph{\empty}As with cancer, initial explanations for overflow metabolism in single-celled organisms used flux-based methods to argue that the growth rate is the fundamental goal of optimal metabolism \citep{kauffman2003}. Subsequently, \citet{molenaar2009a} pointed out that flux-based methods do not account for all the available empirical evidence at the time and proposed instead that differences in protein expression cost is the primary driving force of the choice between OXPHOS and glycolysis. Several other hypotheses, before and after \citet{molenaar2009a}, were formulated to explain overflow metabolism in single-celled organisms, including oxygen limitation \citep{majewski1990}, molecular crowding \citep{vazquez2008}, and competition for membrane space \citep{szenk2017}. With time however, many of these competing hypotheses have coalesced into the comprehensive framework of ``resource allocation" that treats energy production, biogenesis and the protein expression cost of all metabolic pathways as distinct aspects of cellular metabolism that must be optimised together. By decomposing all aspects of cellular metabolism into these three key components-nutrients, bioenergy and machinery-cellular metabolic states may be understood in terms of the cost-vs-benefit of these three components based on their availability and the demand for each of these components \citep{baghdassarian2024}.

\paragraph{\empty}The theory of resource allocation at the individual cellular level has been applied to great effect in bacterial systems, elucidating in many cases the causal basis of how bacteria respond to a change in the nutritional environment. Several studies over the years have found that faster growing cells allocate a higher fraction of their protein machinery to active translation \citep{scott2010, basan2015, peebo2015, keren2016, metzl-raz2017, xia2022}, and much of this literature supported the general notion that glycolytic strategies were more efficient for energy production at higher growth rates due to a lower protein cost \citep{scott2010, basan2015, metzl-raz2017}. This serves to encapsulate the core idea that the costs of the protein machinery play a key role in how metabolic pathways are used by the cell to maximise growth rates \citep{scott2011}. Nevertheless, \citet{you2013} showed that in a wild-type \textit{E. coli} strain, a cyclic adenosine monophosphate (cAMP)-based signalling mechanism exists that helps align the metabolic state of the cell with the nature of nutrient limitation in the environment. Specifically, they found that cAMP-driven signalling allowed for preferential upregulation of catabolic proteome components under carbon limitation and anabolic proteome components under nitrogen deprivation. This suggests that growth rate maximisation may not be a universal determinant of metabolic optimisation in microbes. It has further been shown that the scaling of catabolic gene expression with growth rate in bacteria may optimise for fastest growth on some carbon sources but not others \citep{towbin2017}. In this context, it has been suggested that the upregulation of catabolic genes under carbon limitation may be a preparatory response to future fluctuations in resource availability \citep{new2014, basan2018}, and that specific substrates elicit such a response as a function of their ``quality". A recent study supports this idea by showing not only that bacteria grow at different rates on different substrates, but also that this variation can be attributed to a measure of ``nutrient quality" that depends on the corresponding protein cost of each substrate \citep{mukherjee2024a}. It has also been shown that cellular metabolism in bacteria does not always seek to maximise growth rates as the optimum objective, and instead strive for a balance between allocation to translation and its energetic cost \citep{chure2023}.

\paragraph{\empty}While these protein cost-based theories have substantial experimental and theoretical support \citep{noor2016, basan2018}, some evidence also exists for other kinds of growth optimisation underlying microbial metabolic strategy. For instance, while protein costs may be involved in determining nutrient quality \citep{mukherjee2024a}, it has been shown that bacteria show sequential utilisation or co-utilisation of pairs of carbon sources, depending on whether a given carbon sources enters the glycolytic pathway or the Krebs cycle, indicating a potential role for  network topology to determine metabolic strategies \citep{wang2019}.

\paragraph{\empty}Separately, it has also been shown that faster-growing bacteria show a longer lag phase when introduced to a novel environment with a new carbon source, and lower resistance to starvation than slower-growing bacteria \citep{mukherjee2024a}. \citet{basan2020} likewise found that in \textit{E. coli}, faster growth is associated with a longer lag phase when the bacteria are switched from a standard nutrient medium with a known carbon source to an acetate minimal medium. In this case, it was also shown that the longer lag phase stems from the fact that in an acetate minimal medium, the bacteria must reverse the conventional glycolytic pathway to drive gluconeogenesis instead. \citet{schuetz2012} used a very different theoretical framework based on Pareto optimality to show that metabolic flux states across nine different bacteria operated between two competing objectives: optimal growth on a specific carbon source vs easy switching to other carbon sources. This effectively means that for some bacteria, slower growth in a given medium could have other benefits in terms of adjusting to fluctuating conditions. Similar observations have also been made in budding yeast \citep{metzl-raz2017}, lending support to the idea that microbial metabolic strategies might represent choices between ``generalist" vs ``specialist" growth modes \citep{new2014}. This is interesting particularly because it ties the optimality of microbial metabolic states to the environmental context in a very clear manner. This makes it amenable to more explicit dynamical modelling of multi-species assemblages of bacteria with diverse resource use strategies and a range of environmental conditions \citep{bajic2020, wang2021, wang2025}.

\paragraph{\empty}This literature from microbial metabolism has several key insights to offer for the study of cancer metabolism, but there are some outliers to this body of work that are worth mentioning before we draw the discussion to cancer biology. In a detailed and exhaustive study, \citet{noor2010} used the functional classification of all known enzymes to evaluate a large set of hypothetical metabolic pathway configurations against EMP glycolysis, with the aim of exploring all possible alternative paths to the canonical pathway that are biochemically allowed. They discovered several ``shortcuts" that could potentially shunt some intermediates to other parts of the metabolic network in fewer steps, but importantly, they showed that most of the alternate routes also bypass key precursors that are limiting metabolites for biomass accumulation. This suggests that at least some parts of the glycolytic pathway might not be optimised for maximum growth rate or protein cost, but for ensuring optimal biomass production.

\paragraph{\empty}In a completely different context, \citet{Slavov2014} designed a bioreactor setup to maintain a yeast population at a constant rate of exponential growth to examine any metabolic shifts that might occur with changes in the environment. They identify two clear phases of growth-a glucose-driven phase of rapid growth, followed by an ethanol-driven slower growth phase. Using continuous flux measurements of oxygen and carbon dioxide, the authors demonstrate that while the first phase is predominantly oxidative, there is also some glucose flux into aerobic glycolysis and ethanol production in the first phase. This is generally unexpected at constant growth rates, and the authors then measure both gene and protein expression levels in the same system concurrently to show that the downregulation of OXPHOS towards the end of the first phase could actually stem from the need to reduce the production of reactive oxygen species. Their findings could therefore implicate oxidative stress as an additional variable that could be part of the economics of cellular energetics. A recent study has in fact developed a resource allocation framework for microbial growth that also includes the effect of changing ribosomal concentrations on the redox state of the cell with increasing growth rates \citep{flamholz2025}, demonstrating the growing scope of the resource allocation framework.

\paragraph{\empty}Collectively, the resource allocation framework offers a unified theory explaining a plethora of empirically observed ``growth laws" in bacteria. Further, one of its core insights is the prediction that in many conditions, higher glycolysis can be optimal due to minimizing protein expression cost. Interestingly, several lines of experimental evidence suggest that the growth rate is not always the quantity optimized during metabolism; instead other factors like network topology, biomass production, and recovering from lags matter more in multi-nutrient environments. Regardless of what microbes seek to optimise, resource allocation can still be broadly applied in most of these cases to elucidate the underlying determinants of cellular metabolism. This applicability could be of some use in better understanding cancer metabolism, which we will now explore.

\section{A resource allocation view of cancer metabolism}
\paragraph{\empty}The past two decades of research in microbial metabolism have led to resource allocation theory as a sound framework for rationalising cellular metabolic states. It provides a common thread that can be applied across experimental contexts to understand the cost-benefit balance of microbial metabolic states, regardless of whether or not the growth rate is the basic goal of optimal metabolism. It is tempting to consider applying such a framework to cancer metabolism, particularly because several parallels can be identified between cancer and microbial systems in their metabolic behaviour. For instance, \citet{epstein2014} have shown that aerobic glycolysis in cancer can be interpreted in terms of increased ATP demand from membrane pump activity, proliferation and migration. Interestingly, \citet{mukherjee2024a} have previously noted that faster growth in their \textit{E. coli} system was also associated with cell migration-related strategies. \citet{Birsoy2014a} also found that surviving low glucose availability critically depended on OXPHOS even in cancer cells, paralleling the role of slow growth metabolic strategies in bacteria described above. More traditional flux-based models have also indicated that as with bacteria, solvent constraints \citep{shlomi_genome-scale_2011} and glucose availability \citep{vazquez2010a} could play influential roles in the balance between OXPHOS and glycolysis in cancer cells. The fact that there is increased demand for ribosomal biomass with growth rate in cancerous tissue relative to normal tissue \citep{elser_biological_2007} could likewise point to a microbial-like growth law for cancers that relates growth rate to protein machinery. It is likewise interesting to note that resource sharing phenotypes have been documented widely in both cancer \citep{sonveaux2008, feron2009, chang2015} and microbial systems \citep{wintermute2010}. Several other features of cancer growth and metabolism could parallel microbial metabolism \citep{heiden2011, mayers_famine_2015}, but the application of a comprehensive framework like resource allocation to understand the overall context of these diverse metabolic states still remains limited.

\paragraph{\empty}It must also be noted however, that the application of resource allocation to cancer is complicated by somatic mutations, which can alter the topology of metabolic networks in irreversible and/or tissue-specific manners \citep{yuneva_metabolic_2012}, and provide simpler explanations for a given metabolic state. Studies of metabolism in cancer, and more generally, mammalian metabolism, have also been impeded by the fact that the native biochemical environment of mammalian and cancerous cells is vastly different from in vitro culture conditions, and it is only recently that the implications of these differences are being investigated \citep{muir2017, cantor2017}. Remarkable evidence has also shown that serine biosynthesis pathways are particularly enriched in lung metastases but not the primary tumour in breast cancer \citep{rinaldi2021}. This could potentially be attributed to environmental pyruvate availability in the lung microenvironment vs the primary site, and highlights the importance of the external resource supply in determining which metabolic strategies are optimal. Finally, empirical evidence from a wide range of immortalised mammalian cell lines has shown that the major role of OXPHOS in these cells is to provide precursors for aspartate synthesis rather than ATP production \citep{Sullivan2015}. As with \citet{noor2010} above, this could indicate that the optimality goals of cancer metabolism may not be as straightforward as maximising the specific growth rate. It is therefore an appropriate time to begin asking some fundamental questions in cancer metabolism-are there basic growth laws that relate growth rate to other aspects of cellular energetics? If so, what are the goals of this optimisation problem? How do these goals shift with environmental and/or biological context? Answers to these basic questions would provide the starting point for a resource allocation-based understanding of how cancer metabolism works.

\begin{figure}[ht]
    \centering
    \includegraphics[width=0.8\textwidth, height=0.45\textwidth]{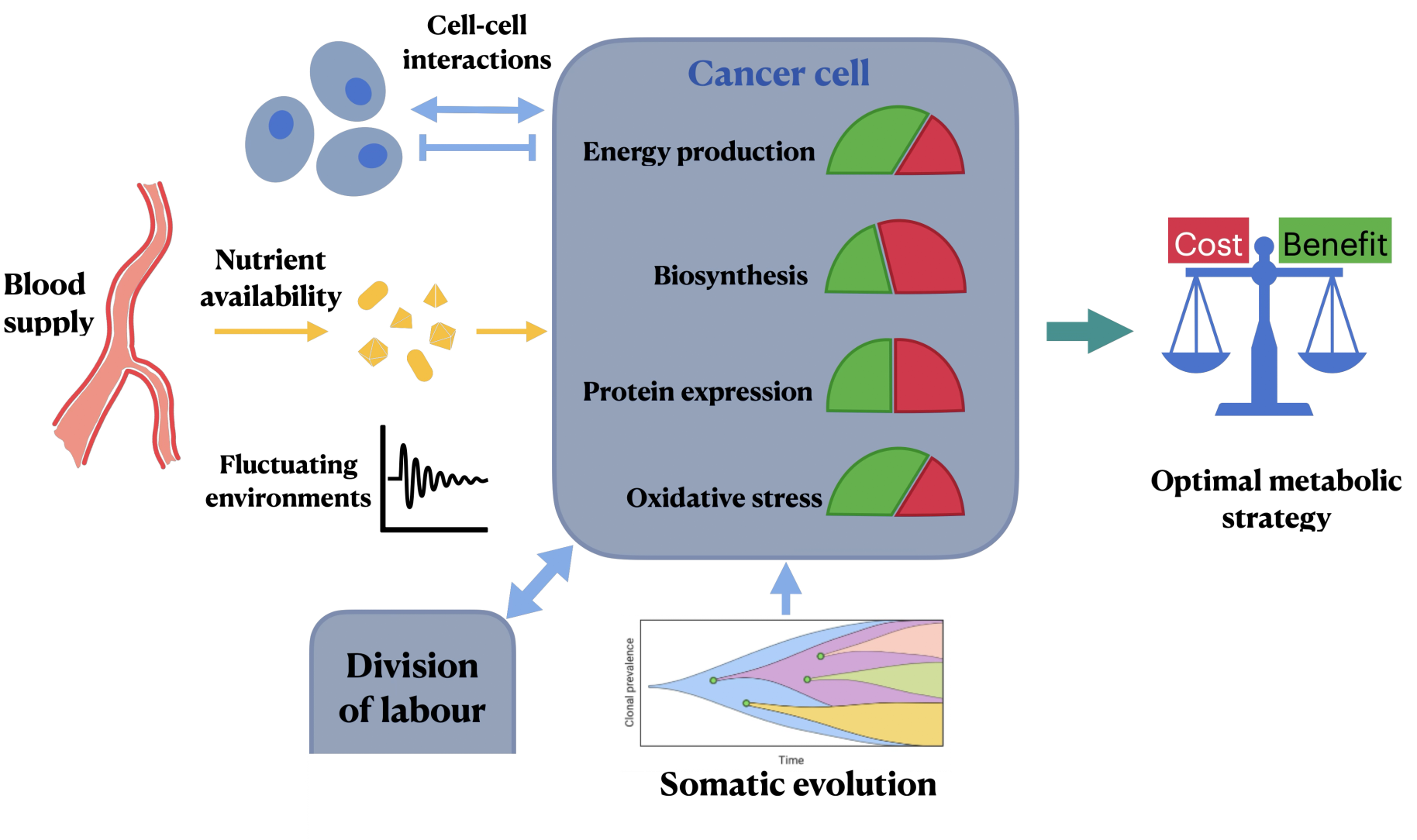}
    \caption{\textbf{Resource allocation as a rational basis for cancer metabolism.} Cancer cells must balance several competing internal processes including energy production (ATP generation), building biomass, protein expression related to specific cell states, and maintaining overall redox balance to prevent oxidative stress. Each of these processes entails particular costs and benefits. This balance is also subject to a range of external factors, including competition or facilitation between cancer and noncancerous cells, supply and availability of nutrients (shown here in yellow), fluctuating nutrient concentrations, pre-existing constraints of multicellular organisation, and \textit{de novo} somatic evolution. Resource allocation posits that the metabolic phenotype of cancer can be understood as an emergent property of this complex optimisation problem.}
    \label{fig1}
\end{figure}

\paragraph{\empty}Despite the many interesting parallels between cancer metabolism and microbial metabolism, cancer is still essentially a phenomenon of multicellularity and it is therefore not obvious that resource allocation principles can be applied to cancer exactly as they have been to microbes. In a compelling and thorough review, \citet{baghdassarian2024} present a broad review of the application of a resource allocation perspective to mammalian cell systems, specifically in the context of multicellular physiology. The extension of resource allocation to multicellular systems is built on the fact that a cell with multiple objectives cannot maximise its gains on all of them simultaneously, as illustrated effectively by the idea of Pareto optimality \citep{shoval_evolutionary_2012, dekel2005, Hausser2019}. For a free-living individual cell, this optimality landscape consists of all possible optimisation objectives that the cell could potentially achieve. In a multicellular context, \citet{baghdassarian2024} argue that division of labour within a tissue allows for this optimisation to occur at the level of the whole tissue rather than that of a single cell. Such an understanding of multicellular organisation provides a firm basis on which cancer phenotypes can be evaluated, as the framework provides a clear way to delineate which aspects of cancer metabolism stem from the organisation of the tissue of origin, and which aspects have developed \textit{de novo}, either through somatic evolution or phenotypic plasticity. Moreover, due to the way it evaluates costs and benefits in terms of bioenergetics, nutrients and machinery, resource allocation theory couples the external environment to the internal state of the cell, thus allowing for metabolic states to be evaluated explicitly in the context in which they would be relevant. Moreover, there are now systems biology platforms that combine dynamical modelling with resource allocation theories, which could allow us to predict the temporal evolution of a range of metabolic strategies \citep{bajic2020}. In this context, it may also be relevant to note that there is considerable evidence for resource sharing interactions between different cancerous and noncancerous subpopulations within a growing tumour \citep{sonveaux2008, feron2009, chang2015}. Dynamical systems modelling is particularly well-suited to studying such interactions to evaluate competing metabolic phenotypes in a complex system, almost like an ecological community \citep{kotler_cancer_2020}. This holds immense potential to expand the current understanding of cancer progression and to identify unique vulnerabilities for effective therapy. Figure \ref{fig1} presents a graphical summary of how a resource allocation-based view of cancer could be devised.

\paragraph{\empty}Here, it is worth mentioning that the role of oxidative stress as a major regulator of cellular metabolic strategy is somewhat unclear. Some studies have found that production of reactive oxygen species (ROS) increases the propensity for cancer cell migration and metastasis \citep{ishikawa2008, porporato2014} while other studies have found that cancer cells in transit experience oxidative stress which impairs metastatic ability \citep{piskounova2015}. As pointed out earlier, oxidative stress has been suggested to be a potential cost of OXPHOS in yeast \citep{Slavov2014}, but a resource allocation-based perspective could potentially evaluate the cost-vs-benefit of both ROS themselves and the protein machinery required for their control, and help delineate the precise role of oxidative stress in cancer growth and progression, as well as microbial growth.

\paragraph{\empty}Finally, an aspect of metabolism potentially unique to cancer is the presence of hybrid metabolic states. Studies have shown that some cancer cells adopt such phenotypes where both OXPHOS and glycolysis are activated, either as an emergent property of how key regulatory genes interact with each other \citep{yu2017, jia2019a} or as a result of the kinetics of enzymatic reactions involved \citep{li2020}. Hybrid states of this kind have rarely been considered explicitly in the resource allocation framework, but they are thought to provide metabolic flexibility to adapt to changing conditions rapidly \citep{yu2017} or improved resistance to radiotherapy \citep{rai2025}. Recent studies have also identified hybrid metabolic states in cancer that can form the basis of drug resistance \citep{jia2020} or represent distinct levels of balance between catabolic and anabolic pathway activity \citep{villela-castrejon2025}. This makes hybrid metabolic states an important consideration for cancer growth, and a resource allocation-based view could help reconcile their emergence and stabilisation in a cancer context. This is even more relevant in the context of fluctuating environments seen in cancers, as environmental variability could be a key factor in promoting such metabolic plasticity and glycolytic activity has been shown to be associated with oscillations in the levels of extracellular metabolites \citep{amemiya2023}. Tumour microenvironments can be highly variable both in space and time \citep{sorg2008, hardee2009}, and this variability has consequences for cancer growth and therapy \citep{martinive2006, dewhirst2007}. The resource allocation framework could potentially help with integrating such unique observations into a rational explanation for cancer metabolism.

\section{Conclusion}
\paragraph{\empty}Despite the consistent prevalence of many metabolic disruptions across cancers, therapeutic intervention on specific aspects of cancer metabolism have so far yielded relatively few successes \citep{luengo_targeting_2017}. Nevertheless, the efficacy of drugs like biguanides as well as that of radiotherapy is clearly impacted by the metabolic state of the tumour \citep{martinive2006, Birsoy2014a, rai2025}, which makes it important to understand the mechanistic basis of cancer metabolic phenotypes. Our review of mammalian and microbial metabolism has revealed several parallels in metabolic regulation between normal mammalian physiology, microbial systems and cancer. The literature suggests that while some aspects of cancer metabolism can be understood in terms of pre-existing regulatory mechanisms in normal physiology, several metabolic features may also be unique to cancer systems. In this context, we show that the framework of resource allocation has been used effectively to investigate the mechanistic basis of metabolism in a range of microbial contexts. We argue that the breadth of its application indicates that the framework of resource allocation could elucidate many distinctive metabolic states in cancer in terms of their relative benefits to cancer cells. This could in turn provide a powerful platform to evaluate competing hypotheses, leading to a more rationalised understanding of how metabolic strategies emerge and evolve in cancer.

\bibliographystyle{plainnat}
\bibliography{main}

\end{document}